\documentclass{jfm}
\usepackage{newtxtext}
\usepackage{newtxmath}
\usepackage{natbib}
\usepackage{hyperref}
\hypersetup{
    colorlinks = true,
    urlcolor   = blue,
    citecolor  = black,
}

\newcommand{\RomanNumeralCaps}[1]
\linenumbers
\usepackage{graphicx} 

\newcommand{\void}{\varepsilon}
\newcommand{\Db}{D_{\textrm{b}}}
\newcommand{\Ar}[1][]{Ar_{*#1}}

\newcommand{\Prb}{Pr_{\textrm{b}}}
\newcommand{\Peb}[1][]{Pe_{\textrm{b}#1}}

\title{Self-similar solution for laminar bubbly flow evolving from a vertical plate.}

\author{N.Valle\corresp{\email{n.vallemarchante@tudelf.nl}},
        \and
        J.W. Haverkort}
\affiliation{       
    Process \& Energy Department,
    Delft University of Technology,
    Leegwaterstraat 39,
    2628 CB Delft,
    The Netherlands}

\begin{document}

\maketitle
\begin{abstract}
    The development of a bubble plume from a vertical gas-evolving electrode is driven by buoyancy and hydrodynamic bubble dispersion. This canonical fluid mechanics problem is relevant for both thermal and electrochemical processes. We adopt a mixture model formulation for the two-phase flow, considering variable density (beyond Boussinesq), viscosity and hydrodynamic bubble dispersion. Introducing a new change of coordinates, inspired by the Lees-Dorodnitsyn transformation, we obtain a new self-similar solution for the laminar boundary layer equations.  The results predict a wall gas fraction and gas plume thickness that increase with height to the power of 1/5 before asymptotically reaching unity and scaling with height to the power 2/5, respectively. The vertical velocity scales with height to the power of 3/5. Our analysis shows that self-similarity is only possible if gas conservation is entirely formulated in terms of the gas-specific volume instead of the gas fraction.    
\end{abstract}

\section{Introduction}
Gas evolution from a vertical plate arises in boiling as well as various electrochemical processes, including water electrolysis for the production of green hydrogen~\citep{Lee2022, LeBideau2020, Khalighi2023}, the production of chlorine and chlorate~\citep{Hedenstedt2017}, or aluminium~\citep{Suzdaltsev2021}. Buoyant forces set the electrolyte in an inexpensive, convective motion, which is advantageous for the mass transport of reactants and products and the removal of heat and bubbles. Similar to heat, bubbles diffuse away from high concentrations. Besides buoyancy, this is the dominant contribution to the bubble motion at low gas fractions~\citep{Dahlkild2001, Schillings2015, rajora2023analytical}. While such two-phase systems resemble thermally-driven natural convection, the analogy of gas fraction with temperature does not work exactly at higher gas fractions due to the increasing importance of other sources of bubble slip. Also, the usual Boussinesq approximation no longer holds due to the strong dependence of density and viscosity with void fraction. This makes reusing well-known self-similarity results for thermal convection~\citep{Ostrach1953,Sparrow1956,Sparrow1959} less accurate unless the gas fraction is very low. Instead, we need to employ a two-fluid formulation to account for the effects of variable density and viscosity~\citep{Ishii2011}. The mixture model is an affordable and common choice for this kind of problem, which has already produced good results in water electrolysis~\citep{Dahlkild2001,Schillings2015,schillings2017four}. An alternative analogy is with particle-laden flows~\citep{Osiptsov1981}, which is only present for forced flows and diluted suspensions. Due to the nonlinearity of the Navier-Stokes equations, buoyant convective motion can develop instabilities -and eventually turbulence- along the flat plate~\citep{Osiptsov1981,Boronin2008}.

The mathematical formulation of the mixture model resembles that of high-speed compressible flow, for which extensive literature on compressible boundary layers exists~\citep{Schlichting1987,Andersson2006} for aerodynamic applications. Whereas gravity is typically disregarded in high-speed flows, it will be essential for gas evolution applications. Furthermore, whereas often a constant wall temperature is assumed, for gas evolution a constant gas flux is usually a more relevant boundary condition. Including buoyancy, taking a constant gas flux at the wall, and considering variable physical properties constitute the main novelties of the analysis presented in this paper. They allow us to obtain new analytical solutions for the industrially important configuration of gas evolution from vertical walls or electrodes. 

\section{Formulation}

\begin{figure}
    \centering
    \includegraphics[width=0.55\columnwidth]{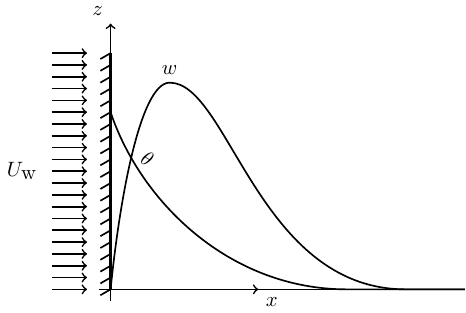}
    \caption{Schematic of the setup for a gas-evolving vertical plate. Typical profiles of vertical velocity $w$ and the gas-to-liquid volume ratio $\theta \propto \void/(1-\void)$ at an arbitrary location.}
    \label{fig:schematic}
\end{figure}
We consider a steady-state two-phase flow of dispersed bubbles and use a two-fluid formulation~\citep{Ishii2011} to describe the average flow field and phase distribution along the vertical plate. Assuming no acceleration between liquid (l) and gas (g) phases, typically valid for bubbly flows, we adopt the mixture model~\citep{Taivassalo1996, Ishii2011}. This model considers the gas-liquid mixture as a single fluid and includes additional closure relations for the unresolved flow features. The mixture density ($\rho$) is the volume-averaged density of the liquid and gas phases, which in terms of the gas volume fraction ($\void$) and liquid density ($\rho_\textrm{l}$) reads:
\begin{equation}
    \label{eqn:rho}
    \rho = \rho_{\textrm{l}} (1-\void)
\end{equation}
where we neglected the contribution of the gas phase to the mixture density due to its typically much lower density (i.e.: $\rho_\textrm{g} \ll \rho_\textrm{l}$). We then introduce the volume- ($\boldsymbol{U}$) and mass- ($\boldsymbol{u}$) averaged velocities.
\begin{align}
    \label{eqn:U}
    \boldsymbol{U} &\equiv \boldsymbol{U}_{\textrm{l}} + \boldsymbol{U}_{\textrm{g}} \\
    \label{eqn:u}
    \rho \boldsymbol{u} & \equiv \rho_{\textrm{l}} \boldsymbol{U}_{\textrm{l}} + \rho_{\textrm{g}} \boldsymbol{U}_{\textrm{g}}
\end{align}
where $\boldsymbol{U}_{\textrm{l}} \equiv (1-\void)\boldsymbol{u}_{\textrm{l}} $ and $ \boldsymbol{U}_{\textrm{g}} \equiv \void \boldsymbol{u}_{\textrm{g}}$ are the superficial liquid and gas velocities expressed in terms of the interstitial velocities $\boldsymbol{u}_{\textrm{l}}$ and $\boldsymbol{u}_{\textrm{g}}$. The superficial slip velocity
\begin{align}
    \label{eqn:j_slip}
    \boldsymbol{U}_{\textrm{slip}} \equiv \varepsilon \left( 1- \varepsilon \right) \left( \boldsymbol{u}_{\textrm{g}} - \boldsymbol{u}_{\textrm{l}} \right)= \void \left( \boldsymbol{u}_{\textrm{g}} -  \boldsymbol{U} \right)
\end{align}
is sometimes referred to as the drift flux or slip flux, as it describes the relative volume flux of the gas with respect to the mixture, and can be seen as a sub-scale model. From the previous definitions, we obtain the following relation between mass- and volume-averaged velocities
\begin{equation}
    \boldsymbol{u} = \boldsymbol{U} -
    \frac{\boldsymbol{U}_{\textrm{slip}}}{1-\void}
    \label{eqn:u-j-jslip}
\end{equation}

Central to the mixture model are the semi-empirical relations that describe the slip velocities. Slip velocities result from the balance between drag and bubble forces, like lift, buoyancy or bubble-bubble interactions. In defining the slip velocities, we assume that the vertical buoyant rise velocity of bubbles relative to the liquid can be neglected with respect to the liquid velocity itself. In the horizontal direction, we neglect the lift forces and consider only the bubble-bubble interactions that lead to hydrodynamic dispersion, which we model using
\begin{align}
    \label{eqn:j_hd}
    \boldsymbol{U}_{\textrm{slip}}=\boldsymbol{U}_{\textrm{Hd}}=
        -\frac{D_{\textrm{b}}}{\left( 1-\void \right)^{2}} \nabla \void 
\end{align}
Equations \eqref{eqn:j_slip} and~\eqref{eqn:j_hd} then give the volumetric gas flux as the sum of an advective and diffusive flux: $\boldsymbol{U}_{\textrm{g}}=\boldsymbol{U}\varepsilon - \frac{D_{\textrm{b}}}{\left( 1-\void \right)^{2}} \nabla \varepsilon  $ with a diffusion coefficient equal to $\frac{D_{\textrm{b}}}{\left( 1-\void \right)^{2}}  $. The hydrodynamic dispersion coefficient $D_{\textrm{b}}$ was for low solid particle volume fractions found to be approximately given by the particle radius times the Stokes settling velocity~\citep{Ham1988,Nicolai1995}, so
\begin{equation}
D_{\textrm{b}} \approx \frac{d_{\textrm{b}} w_\textrm{St}}{2}   \label{eqn:Db}
\end{equation}
with $d_{\textrm{b}}$ the bubble diameter and $w_\textrm{St} = \frac{ g d_{\textrm{b}}^2}{18\nu_{\textrm{l}}}$ the Stokes rise velocity in terms of the kinematic liquid viscosity $\nu_{\textrm{l}}=\mu_{\textrm{l}}/\rho_{\textrm{l}}$. In Eq.~\eqref{eqn:j_hd} we divided by an \emph{ad-hoc} additional factor $1/\left( 1-\void \right)^{2}$ to approximately account for the additional repulsive effects that arise as the gas fraction increases. This can be seen as one of the mechanisms keeping volume fractions below $1$. For particle suspensions, this is often modelled by adding a `solid pressure'~\citep{Johnson1987}, which has a mathematically similar effect. \citet{rajora2023analytical} included a similar term with a lower power of 1 instead of 2 and an arbitrary maximum gas fraction. While this form of the slip velocity expression is arguably the least validated assumption of our model, it does reduce to the correct empirically validated limit at low gas fractions and allows us to describe the essence of solid pressure effects in a simplified way. Only this particular form allows for a self-similarity solution of the boundary layer equations, as is demonstrated in the Supplementary Material. 

For flows developing a laminar boundary layer along a vertical plate, we modify the compressible version of Prandtl's boundary layer theory~\citep{Andersson2006} to include gravity. 
\begin{align}
    \label{eqn:mass}
        \partial_x \left( \rho u \right)
    +   \partial_z \left( \rho w \right)
    &=  0  \\
    \label{eqn:momentum}
        \rho u w_x
    +   \rho w w_z
    &=  \partial_x \left(\mu w_x \right)
    +   \left( \rho_\textrm{l} - \rho \right) g \\
    \label{eqn:void}
        \rho u \partial_x \frac{1}{\rho}
    +   \rho w \partial_z \frac{1}{\rho}
    &=  \partial_x \left(\rho_\textrm{l}\frac{D_{\textrm{b}} }{1-\void} \partial_x \frac{1}{\rho} \right)
\end{align}
where $u$ and $w$ are the components of $\boldsymbol{u}$ in the horizontal $x$ and vertical $z$-direction, respectively. Far away from the wall, where $\left(\rho_{\textrm{l}} - \rho \right)g=0$, a hydrostatic pressure gradient is assumed to cancel the gravitation force. Close to the wall, this term is positive, due to the upwards force buoyancy exerts on the mixture. The gas transport equation~\eqref{eqn:void} follows from taking the divergence of~\eqref{eqn:u-j-jslip}, $\nabla\cdot\boldsymbol{U}=0$ (since gas and liquid are conserved, i.e.: $\nabla\cdot\boldsymbol{U}_{\textrm{g}}=\nabla\cdot\boldsymbol{U}_{\textrm{l}}=0$), inserting ~\eqref{eqn:rho} and~\eqref{eqn:j_hd}, and finally invoking $1/\left( 1-\void \right)^2 \partial_x \void = \partial_x \left( 1 / \left( 1-\void \right) \right)$. Note that additional $\rho$ and $1/\rho$ terms are included in the convective term to render the final form in terms of the mass-averaged mixture velocity $(u,w)$ and the specific volume $1/\rho$.

The resulting equations~\eqref{eqn:mass}-\eqref{eqn:void} resemble that of high-speed aerodynamics, inspiring our solution approach. It is worth mentioning that the last equation corresponds to the conservation of the gas phase expressed in terms of specific volume ($1/\rho$).

We assume the following rheological relation for mixture viscosity~\citep{Ishii1979}:
\begin{equation}  \label{eqn:mu}
    \mu = \frac{\mu_{\textrm{l}}}{1-\void}
\end{equation}
and define the following boundary conditions:
\begin{align}
    \label{eqn:bc-vel}
    u\rvert_{x=0} = w\rvert_{x=0} &= 0 , &
    \lim_{x\to\infty}w &= 0 \\
    \label{eqn:bc-gas}
    \left.-\frac{D_{\textrm{b}}}{\left( 1-\void \right)^2} \frac{\partial \void}{\partial x}\right\vert_{x=0} &= 
        U_{\textrm{w}} (1-\void_{\textrm{w}}) , &
    \lim_{~x\to \infty} \void &= 0
\end{align}
which completes the formulation of the problem. The Neumann boundary condition for the gas fraction follows from equations \eqref{eqn:j_slip} and \eqref{eqn:j_hd}, which give $ - \frac{D_{\textrm{b}}}{\left( 1-\void \right)^{2}} \nabla \varepsilon = \boldsymbol{U}_{\textrm{g}}-\boldsymbol{U}\varepsilon$, and assuming $\boldsymbol{U}_{\textrm{l}}\approx0$ owing to the very low specific volume of the liquid with respect to the gas phase. In typical electrochemical systems, the superficial gas velocity at the wall $U_{\textrm{w}} = j\mathcal{V}_{\textrm{m}}/n\textrm{F}$ can be related to the current density $j$, the number of gas molecules $n$ produced per electron converted in the reaction, Faraday's constant $\textrm{F}$, and the molar volume of the gas $\mathcal{V}_\textrm{m}$ --typically given by the ideal gas law as $\mathcal{V}_{\textrm{m}} = \textrm{R}T/p$, where $\textrm{R}$ is the ideal gas constant, $T$ the absolute temperature and $p$ the pressure. 

\section{Methodology}

Next, we summarize the steps leading to the development of the self-similarity solution. The procedure is similar to that employed in obtaining a self-similar solution for high-speed compressible flows~\citep{Andersson2006}, with variations to include the dependency on gravity. 

\emph{Stream function formulation.} We first introduce a stream function $\Psi$ for the mass flux
\begin{align}
    \label{eqn:streamfunction}
    \rho u &= -\Psi_z ,   &   \rho w &= \Psi_x
\end{align}
which makes equation~\eqref{eqn:mass} automatically satisfied. We rewrite the other governing equations as:
\begin{align}
    \label{eqn:momentum-stream}
    -   \Psi_z \partial_x \frac{\Psi_x}{\rho}
    +   \Psi_x \partial_z \frac{\Psi_x}{\rho}
    &=  \partial_x \left( \frac{\rho_{\textrm{l}}\mu_{\textrm{l}}}{\rho} \partial_x \frac{\Psi_x}{\rho} \right)
    +   \left( \rho_{\textrm{l}}-\rho \right)g\\
    \label{eqn:void-stream}
    -   \Psi_z \partial_x \frac{1}{\rho}
    +   \Psi_x \partial_z \frac{1}{\rho}
    &=  \partial_x \left(D_{\textrm{b}}\frac{ \rho_\textrm{l}^2}{\rho} 
\partial_x \frac{1}{\rho} \right)
\end{align}
where subscripts $x$ and $z$ denote partial derivatives.

\emph{Dorodnitsyn transformation.} Now, we introduce the following self-similarity variable $\eta$
\begin{align}
    \label{eqn:transformation}
    \eta = \frac{\Ar[z]^{1/5}}{z}\int_0^x \frac{\rho}{\rho_{\textrm{l}}} dx'
\end{align}
where we introduce the modified Archimedes number $\Ar[z] = gU_\textrm{w}z^4/\nu_\textrm{l}^2 D_\textrm{b}$. This self-similarity transformation can be seen as combining the transformation proposed by \citet{Sparrow1956} for Boussinesq flows developing along a vertical flat plate subject to constant heat flux with the original Dorodnitsyn transformation~\citep{Andersson2006} for compressible flows. Note that in this particular type of transformation, the new coordinate $\eta$ explicitly depends on the vertical coordinate $z$, while the dependency on the horizontal $x$ coordinate is introduced via the density-weighted integral. The intent of such transformation is to simplify the calculation of $w$ in Prandtl's boundary layer equations. In particular: 
\begin{equation}
    w = \frac{\Psi_x}{\rho} = \frac{\eta_x}{\rho}\Psi_\eta = \frac{\Ar[z]^{1/5}}{z \rho_\textrm{l}} \Psi_\eta    \label{eqn:w}
\end{equation}
which removes the explicit dependency on $\rho$ of the terms involving $w$, arguably simplifying the treatment of the boundary layer equations despite the variable density. Step-by-step details on the application of such transformation can be found in the Supplementary Material. 

\emph{Separation of variables.}
Next, we adopt the following transformation for the unknowns $\Psi$ and $1/(1-\void)$ in terms of the self-similar functions $f(\eta)$ and $\theta(\eta)$
\begin{align}
    \label{eqn:Psi-void-transformation}
    \Psi &= \mu_{\textrm{l}} \Ar[z]^{1/5} f(\eta)\\
    \label{eqn:void2}
    \frac{1}{1-\void} &= 1 +\Peb[z] \Ar[z]^{-1/5} \theta(\eta)
\end{align}
Where we introduce the P\'{e}clet number $\Peb[z] = U_\textrm{w} z / D_\textrm{b}$. These transform equations~\eqref{eqn:momentum-stream} and~\eqref{eqn:void-stream} into the system
\begin{align}
    \label{eqn:self-similar-f}
    f''' &= \frac{3}{5} f'^2 - \frac{4}{5} f f'' - \theta \\
    \label{eqn:self-similar-theta}
    \theta'' &= \frac{\Prb}{5} \left(f'\theta - 4 f\theta'\right) 
\end{align}
where the bubble Prandtl number $\Prb = \nu_{\textrm{l}} / D_{\textrm{b}}$.
The system is subject to the boundary conditions
\begin{align}
    \label{eqn:self-similar-bc-f}
    f\rvert_{\eta=0} = f'\rvert_{\eta=0} &=0 & \lim_{\eta\to \infty}f'     &=0\\
    \label{eqn:self-similar-bc-theta}
    \theta'\rvert_{\eta=0} &= -1              & \lim_{\eta\to \infty}\theta &=0
\end{align}
where a prime ($'$), denotes a derivative with respect to $\eta$ to shorten the expressions. These transformations now allow us to derive a similarity solution. The system of equations~\eqref{eqn:self-similar-bc-f}-\eqref{eqn:self-similar-theta} is the equivalent as the classical result for Boussinesq natural convection with constant heat flux~\citep{Sparrow1956}. 
However, the most remarkable change is the shape of the buoyant scalar (i.e., $\void$). In particular, equation~\eqref{eqn:void} gives
\begin{equation}
    \label{eqn:self-similar-scalar-dorodnitsyn}
    \void = \frac{\Peb[z] \Ar[z]^{-1/5}\theta}{1+\Peb[z] \Ar[z]^{-1/5}\theta}
\end{equation}
whereas in the thermal convection case~\citep{Sparrow1956}   $   \void = \Peb[z] \Ar[z]^{-1/5}\theta$, which agrees with equation~\eqref{eqn:self-similar-scalar-dorodnitsyn} when $\Peb[z] \Ar[z]^{-1/5}\theta \approx 0$. This is expected since for low gas fractions, we should recover the Boussinesq hypothesis. The denominator in equation~\eqref{eqn:self-similar-scalar-dorodnitsyn} ensures that the gas fraction does not exceed unity, a phenomenon in bubbly flow with no equivalent in thermal natural convection.

The vertical velocity field can be recovered from~\eqref{eqn:w} and~\eqref{eqn:Psi-void-transformation} as
\begin{align}
    \label{eqn:w-vs-z}
    w &= \frac{\nu_\textrm{l}}{z} \Ar[z]^{2/5} f'
\end{align}
which gives a scaling with height proportional to $ z^{3/5}$, equal to the thermal convection solution~\citep{Sparrow1956}.

The mass flow rate per unit width [kg/m/s] is then equal to:
\begin{equation}  \label{eqn:mflux}
    \int_0^\infty \rho w dx = \int_0^\infty \Psi_x dx = \Psi_\infty = \mu_{\textrm{l}}\Ar[z]^{1/5}f_\infty
\end{equation}
which scales with $z^{4/5}$, accounting for the suction in the horizontal direction from outside the plume.

The wall shear stress $\tau_\textrm{w}$ [Pa] is given by
\begin{equation}
    \label{eqn:tau_w-vs-z}
    \tau_\textrm{w} \equiv 
    \left. { \mu \frac{\partial w}{\partial x}}\right\vert_\textrm{w} = \\
        \mu_\textrm{l} \frac{\nu_\textrm{l}}{z^2} \Ar[z]^{3/5} f''_\textrm{w}
\end{equation}
where we used equation~\eqref{eqn:mu} for the mixture viscosity $\mu = \frac{\mu_{\textrm{l}}}{1- \varepsilon}$ as well as equations~\eqref{eqn:rho}, \eqref{eqn:transformation}, and~\eqref{eqn:w-vs-z}. 

Interestingly, the wall shear stress continues to show the same scaling with height and current density also when the wall gas fraction asymptotically tends to unity. However, the wall shear rate $w'_{\textrm{w}} = \frac{\tau_{\textrm{w}} }{  \mu} = \frac{\tau_{\textrm{w}} }{  \mu_{\textrm{l}}} \left( 1- \varepsilon_{\textrm{w}} \right) $ [s$^{-1}$] does show a transition between $z^{2/5}$ and $z^{3/5}$. Using equation~\eqref{eqn:self-similar-scalar-dorodnitsyn} this gives
\begin{equation}  \label{eqn:wsr}
w'_{\textrm{w}} =  \frac{\nu_\textrm{l}}{z^2} \frac{f''_\textrm{w} \Ar[z]^{3/5}}{1+\Peb[z] \Ar[z]^{-1/5}\theta_{\textrm{w}} }
\end{equation}
Finally, we may define the dimensionless gas plume thickness as $\frac{\delta_{\textrm{g}}}{z} = - \frac{z^{-1}}{\partial \eta / \partial x } \frac{ \theta_{\textrm{w}} }{\theta'_{\textrm{w}} } $. It follows from the boundary conditions in equation~\eqref{eqn:self-similar-bc-theta} and equations~\eqref{eqn:transformation} and~\eqref{eqn:rho} that
\begin{equation}  \label{eqn:dldelta}
\frac{\delta_{\textrm{g}}}{z} =  \Ar[z]^{-1/5} \left( 1 + \Peb[z]\Ar[z]^{-1/5} \theta_{\textrm{w}} \right) \theta_{\textrm{w}}
\end{equation} 
which shows a transitional scaling between $z^{-4/5}$ and $z^{-3/5}$.

\section{Results}
The final system of equations~\eqref{eqn:self-similar-f}-\eqref{eqn:self-similar-bc-theta} can be solved by numerical integration for different values of $\Prb$, which we choose corresponding to a wide range of bubble diameters.

\begin{figure*}
    \begin{centering}
    \includegraphics[width=0.49\columnwidth]{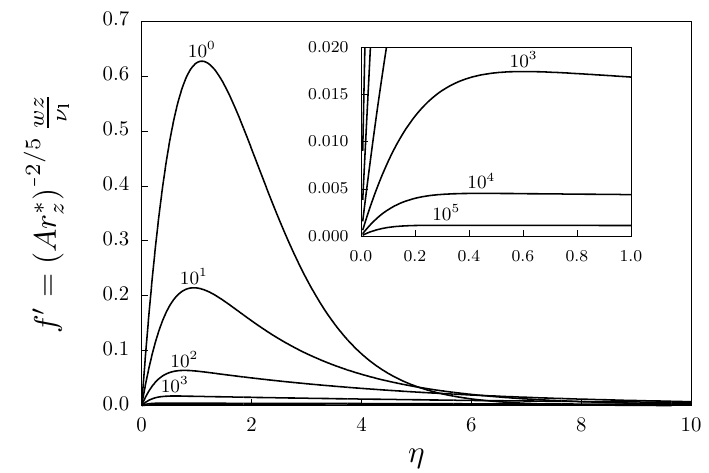}
    \includegraphics[width=0.49\columnwidth]{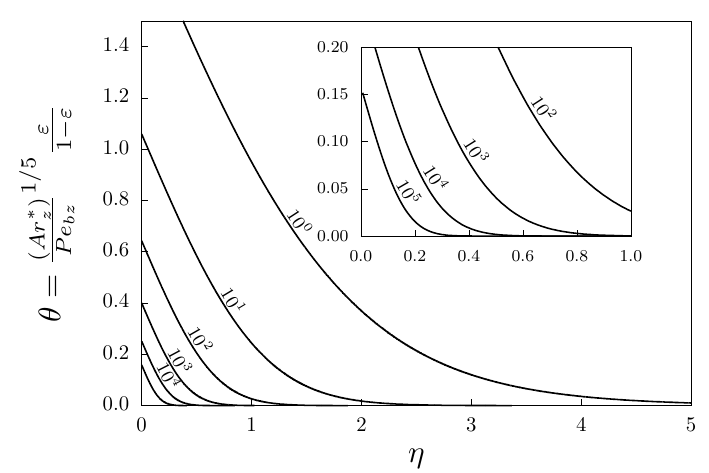}
                \caption{The numerical solution of equations~\eqref{eqn:self-similar-f} and~\eqref{eqn:self-similar-theta} for $f'$ (left) and $\theta$ (right) near the wall for several $\Prb$ numbers as indicated in the figure.
            The computational domain is $\eta \in [0,50]$, but for visualization purposes, a shorter interval is shown.}
        \label{fig:res-vs-eta}
    \end{centering}
\end{figure*}
The numerical method consists of a classical shooting algorithm, which adjusts the value for $\theta$ at $\eta=0$ iteratively until the boundary conditions are satisfied. The domain of integration ranges from $\eta=0$ to $\eta = 50$. This range was chosen such that parameters in the vicinity of the wall do not change substantially upon a further increase. For $f_\infty$, convergence was achieved for a substantially larger domain up to $\eta=800$. Results for $f'$ and $\theta$ are shown in Figure~\ref{fig:res-vs-eta}. Through equations~\eqref{eqn:self-similar-scalar-dorodnitsyn} and~\eqref{eqn:w-vs-z} these dimensionless quantities are related to those of the vertical velocity $w$ and gas fraction $\varepsilon$, respectively. 

\begin{figure}
    \centering
    \includegraphics[width=0.85\textwidth]{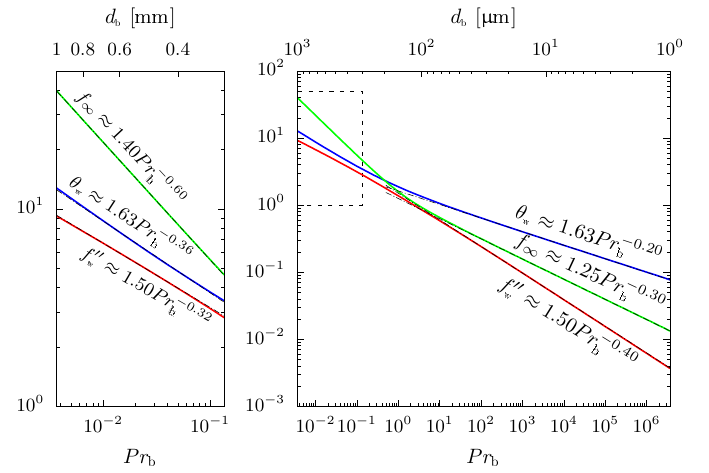}
    \caption{The wall values $\theta_\textrm{w}$ and $f''_{\textrm{w}}$ and $f_{\infty}$ as a function of the bubble Prandtl number $\Prb = \nu / D_{\textrm{b}}$. The upper $x$-axis shows the corresponding bubble diameter using equation~\eqref{eqn:Db}, assuming typical values of $g=9.81$ m/s$^2$ and $\nu_{\textrm{l}}=10^{-6}$ m$^2$/s. We note that for bubbles with $d_{\textrm{b}} \gtrsim 100$ $\upmu$m, $w_{\textrm{St}}$ used in Eq.~\eqref{eqn:Db} no longer corresponds to the buoyant slip velocity as Stokes drag becomes invalid.}
    \label{fig:theta-d-vs-Prb}
\end{figure}

From these profiles, we can obtain relevant parameters such as wall gas fraction $\void_\textrm{w}$ and wall shear stress $\tau_\textrm{w}$, which depend on $\theta_\textrm{w}$ and $f''_\textrm{w}$; and also on vertical mass flow rate, which depends on $f_\infty$. Values of $\theta_\textrm{w}$, $f''_\textrm{w}$ and $f_\infty$ are reported in Figure~\ref{fig:theta-d-vs-Prb} for different $\Prb$ numbers. All profiles show a transition at $\Prb\approx1$, corresponding to the switch in the predominance of the momentum diffusivity $\nu$ over the bubble diffusivity $D_{\textrm{b}}$. For convenience, Figure~\ref{fig:theta-d-vs-Prb} has as an additional axis showing typical bubble sizes assuming Stokes drag and an aqueous-like viscosity. Results show different asymptotic behaviours for small bubbles ($d_{\textrm{b}}<100 ~\upmu\textrm{m}$, typical for electrolytic bubbles) and larger bubbles ($d_{\textrm{b}}>100 ~\upmu\textrm{m}$, typical for bubbles due to boiling). A power-law scaling is obtained for the two regions, shown in the figure,
resulting in the approximate asymptotic relations for both parameters
\begin{align}
    \label{eqn:theta_w-vs-Pr_b}
    \theta_\textrm{w} &\approx \begin{cases} 1.63\Prb^{-1/5}& \Prb \gtrsim 1 \\ 1.63 \Prb^{-1/3} & \Prb \lesssim 1  \end{cases}\\
    \label{eqn:f''_w-vs-Pr_b}
    f''_\textrm{w}    &\approx \begin{cases} 1.50\Prb^{-2/5}& \Prb \gtrsim 1  \\ 1.50 \Prb^{-1/3} & \Prb \lesssim 1 \end{cases}\\
    \label{eqn:f_infty-vs-Pr_b}
    f_\infty    &\approx \begin{cases} 1.40\Prb^{-3/5}& \Prb \gtrsim 1  \\ 1.25 \Prb^{-3/10} & \Prb \lesssim 1 \end{cases}
\end{align}
Approximations combining both asymptotic limits can be found in the Supplementary Material.
These values can be used to approximate wall gas fraction $\void_\textrm{w}$, wall shear stress $\tau_\textrm{w}$, and wall shear rate $w'_{\textrm{w}}$ through equations~\eqref{eqn:self-similar-scalar-dorodnitsyn}, \eqref{eqn:tau_w-vs-z} and~\eqref{eqn:wsr}, respectively, to give
\begin{align}
    \label{eqn:void-vs-z}
    \frac{ \void_\textrm{w} }{1 - \void_\textrm{w}}
        &\approx 1.63 \times
        \begin{cases}
              \left( \frac{ \nu_\textrm{l}  U_{\textrm{w}}^4 z }{g D_\textrm{b}^3  }  \right)^{1/5} 
            & d_{\textrm{b}}\lesssim 100 ~\upmu\textrm{m} \\
             \left( \frac{\nu_\textrm{l}^{1/3}   U_{\textrm{w}}^4 z}{g D_\textrm{b}^{7/3} }  \right)^{1/5} 
 & d_{\textrm{b}}\gtrsim 100 ~\upmu\textrm{m}
        \end{cases}
\end{align}

\begin{equation}
    \label{eqn:shear-vs-z-small-bubbles}
    \tau_\textrm{w} \approx
    1.5 \rho_{\textrm{l}} \times \begin{cases}
         \left(  \frac{g^3 \nu_\textrm{l}^2 U_\textrm{w}^3  z^2   }{  \Db }\right)^{\frac{1}{5}}  & d_{\textrm{b}}\lesssim100 ~\upmu\textrm{m} \\
        \left(  g^3   \left(\nu_\textrm{l}^{7}/\Db^{4} \right)^{\frac{1}{3}}  U_\textrm{w}^3 z^2 \right)^{\frac{1}{5}} & d_{\textrm{b}}\gtrsim100 ~\upmu\textrm{m} \\
    \end{cases}
\end{equation}
and
\begin{equation}
    \label{eqn:w'-vs-z-small-bubbles}
    w'_\textrm{w} \approx
    1.5\times \begin{cases}
         \frac{ \left(  \frac{g^3 U_\textrm{w}^3  z^2  }{  \Db \nu_\textrm{l}^3  }\right)^{1/5} }{1+1.63 \left( \frac{ \nu_\textrm{l}   U_{\textrm{w}}^4 z }{g D_\textrm{b}^3  }  \right)^{1/5} } & d_{\textrm{b}}\lesssim100 ~\upmu\textrm{m} \\
        \frac{\left(  \frac{ g^3 U_\textrm{w}^3 z^2  }{ \left(\nu_\textrm{l}^{8} \Db^{4} \right)^{\frac{1}{3}} } \right)^{1/5}}{1+1.63 \left( \frac{\nu_\textrm{l}^{1/3}  U_{\textrm{w}}^4 z }{g D_\textrm{b}^{7/3} }  \right)^{1/5} } & d_{\textrm{b}}\gtrsim100 ~\upmu\textrm{m} \\
    \end{cases}
\end{equation}
This last relation shows a transition from a proportionality with $U_{\textrm{w}}^{3/5}z^{2/5}$ at low wall gas fractions to $U_{\textrm{w}}^{-1/5}z^{1/5}$ at wall gas fractions close to one. Interestingly, the viscosity increases faster with increasing $U_{\textrm{w}}$ than the wall-shear stress, resulting in a decreasing wall shear rate with increasing gas flux in this regime. This limiting result strongly depends on the exact relation between effective viscosity and gas fraction viz. equation~\eqref{eqn:mu} and the assumed maximum gas fraction of unity.

The mass flux per unit width, from equation~\eqref{eqn:mflux} reads

\begin{equation}  \label{eqn:mflux-dim}
    \int_0^\infty \rho w dx =  \begin{cases}
         1.4\rho_{\textrm{l}} \left( g  D_\textrm{b}^2 U_\textrm{w}z^4 \right)^{1/5}  & d_{\textrm{b}}\lesssim100 ~\upmu\textrm{m} \\
        1.25\rho_{\textrm{l}} \left( g \nu_\textrm{l}^{3/2} D_\textrm{b}^{1/2}  U_\textrm{w}z^4\right)^{1/5}  & d_{\textrm{b}}\gtrsim 100 ~\upmu\textrm{m} 
    \end{cases}
\end{equation}
This shows an almost linear increase with $z$ and a much weaker dependence on the wall gas flux $U_{\textrm{w}}$. \cite{rajora2023analytical} found a linear dependence on both $z$ and $U_{\textrm{w}}$, for a plume in the shape of a step-function. When $ \Peb[z] \Ar[z]^{-1/5} \gg 1$ we have $\varepsilon_{\textrm{w}}\approx 1$ and the gas plume described by equation~\eqref{eqn:self-similar-scalar-dorodnitsyn} similarly becomes a smoothed step-function. However, in this limit, there is no liquid inside the plateau region of the gas plume and the analysis of~\cite{rajora2023analytical} fails in this case.

Finally, using equations~\eqref{eqn:dldelta},~\eqref{eqn:theta_w-vs-Pr_b} and~\eqref{eqn:void-vs-z} the gas plume thickness becomes
\begin{equation}
\delta_{\textrm{g}} \approx 1.63 \left(\frac{z}{ g U_\textrm{w}  } \right)^{\frac{1}{5}} \times \begin{cases} \nu_\textrm{l}^{\frac{1}{5}} D_\textrm{b}^{\frac{2}{5}}\left( 1+\left( \frac{ \nu_\textrm{l}   U_{\textrm{w}}^4 z }{g D_\textrm{b}^3  }  \right)^{1/5} \right)  & d_{\textrm{b}}\lesssim100 ~\upmu\textrm{m} \\ 
\nu_\textrm{l}^{\frac{1}{15}} D_\textrm{b}^{\frac{8}{15}}\left( 1+ \left( \frac{\nu_\textrm{l}^{1/3}  U_{\textrm{w}}^4 z }{g D_\textrm{b}^{7/3} }  \right)^{1/5} \right) & d_{\textrm{b}}\gtrsim100 ~\upmu\textrm{m}
\end{cases}
\end{equation}
This shows a transition from a proportionality with $\left(z / U_{\textrm{w}} \right)^{1/5}$ at low gas fraction to a proportionality with $U_{\textrm{w}}^{3/5}z^{2/5}$ at high wall gas fractions. The latter positive dependence on gas flux is in agreement with most experimental findings in which the plume thickness increases with increasing current density. A similar transition was found by~\citet{rajora2023analytical} using approximate methods, where at high gas fractions, depending on $\Prb$, a proportionality with between $\left( U_{\textrm{w}}z\right)^{1/3}$ and $U_{\textrm{w}}^{0.73}z^{0.43}$ was obtained. The result obtained here is virtually exact, valid for all wall gas fractions, and arguably more elegant. 

\section{Conclusion}
We have developed a new self-similarity solution for laminar bubbly flow evolving from a vertical plate, which considers variable density, viscosity, and hydrodynamic dispersion, all depending on the local gas fraction. Results for both small and large bubbles produce simple power law relations for the wall shear stress and the gas fraction at the wall, which are critical parameters in process technology. The wall strain rate and gas plume thickness show a transition between low to moderate gas fractions and gas fractions near one.

Our theoretical analysis reveals that a self-similarity solution is only possible if the bubble diffusivity is proportional to $(1-\void)^{-2}$, which is equivalent to formulating the gas fraction convection-diffusion equation in terms of specific volumes and using a constant diffusivity. While this is a convenient assumption that allows for a neat analytical treatment, it can also be physically motivated to approximately model increased diffusion at high gas fractions, preventing the system from reaching nonphysical gas fractions $\void > 1$.

By introducing a novel change of coordinates and variables  we observe that the solutions show an asymptotic transition between the classical Boussinesq result for relatively low gas flow rates and short heights and a new asymptotic solution for which the wall gas fraction tends to one.


\backsection[Funding]{We acknowledge the Dutch Research Council (NWO) for funding under grant agreement KICH1.ED04.20.011. }

\backsection[Declaration of interests]{The authors report no conflict of interest.}


\backsection[Author ORCIDs]{N. Valle, https://orcid.org/0000-0003-2140-041X; J.W. Haverkort, https://orcid.org/0000-0001-5028-5292}

\bibliographystyle{jfm}
\bibliography{library.bib, bib2.bib}
\end{document}